\documentclass[reprint,superscriptaddress,amsmath,amssymb,aps,prl,nofootinbib]{revtex4-2}
\usepackage{lipsum} 
\usepackage{soul}
\usepackage{textcomp}  
\usepackage{graphicx}  
\usepackage{hyperref}  
\usepackage{siunitx}
\usepackage{xspace}
\usepackage{tabularx}
\usepackage[symbol]{footmisc}
\usepackage{IEEEtrantools}
\usepackage{xcolor}

\DeclareSIUnit\gauss{G}
\DeclareSIUnit\rad{rad}

\newcommand{\ket}[1]{\ensuremath{\lvert #1 \rangle}\xspace}

\newcommand{\figref}[2][]{Fig.\,\ref{#2}#1}

\makeatletter
\def\maketitle{
	\@author@finish
	\title@column\titleblock@produce
	\suppressfloats[t]}
\makeatother

\begin{document}
	
    \newcommand{\partitle}[1]{\section{#1}}

     \newcommand{\papertitle}{Universal gates for a metastable qubit in strontium-88}
	\title{\papertitle{}}

    \author{Renhao Tao}
	   \affiliation{Max-Planck-Institut f\"{u}r Quantenoptik, 85748 Garching, Germany}
	   \affiliation{Munich Center for Quantum Science and Technology (MCQST), 80799 Munich, Germany}
	   \affiliation{Fakultät für Physik, Ludwig-Maximilians-Universit\"{a}t, 80799 Munich, Germany}

    \author{Ohad Lib}
	   \affiliation{Max-Planck-Institut f\"{u}r Quantenoptik, 85748 Garching, Germany}
	   \affiliation{Munich Center for Quantum Science and Technology (MCQST), 80799 Munich, Germany}

    \author{Flavien Gyger}
	   \affiliation{Max-Planck-Institut f\"{u}r Quantenoptik, 85748 Garching, Germany}
	   \affiliation{Munich Center for Quantum Science and Technology (MCQST), 80799 Munich, Germany}

    \author{Hendrik Timme}
	   \affiliation{Max-Planck-Institut f\"{u}r Quantenoptik, 85748 Garching, Germany}
	   \affiliation{Munich Center for Quantum Science and Technology (MCQST), 80799 Munich, Germany}
	
	\author{Maximilian Ammenwerth}
	  \affiliation{Max-Planck-Institut f\"{u}r Quantenoptik, 85748 Garching, Germany}
	   \affiliation{Munich Center for Quantum Science and Technology (MCQST), 80799 Munich, Germany}

    \author{Immanuel Bloch}
	   \affiliation{Max-Planck-Institut f\"{u}r Quantenoptik, 85748 Garching, Germany}
	   \affiliation{Munich Center for Quantum Science and Technology (MCQST), 80799 Munich, Germany}
	   \affiliation{Fakultät für Physik, Ludwig-Maximilians-Universit\"{a}t, 80799 Munich, Germany}

   \author{Johannes Zeiher}
   \email{johannes.zeiher@mpq.mpg.de}
    \affiliation{Max-Planck-Institut f\"{u}r Quantenoptik, 85748 Garching, Germany}
    \affiliation{Munich Center for Quantum Science and Technology (MCQST), 80799 Munich, Germany}
    \affiliation{Fakultät für Physik, Ludwig-Maximilians-Universit\"{a}t, 80799 Munich, Germany}
	
	\date{\today}
	
	\begin{abstract}
        Metastable atomic qubits are a highly promising platform for the realization of quantum computers, owing to their scalability and the possibility of converting leakage errors to erasure errors mid-circuit.
        Here, we demonstrate and characterize a universal gate set for the metastable fine-structure qubit encoded between the ${^3}\text{P}_0$ and ${^3}\text{P}_2$ states in bosonic strontium-88.
        We find single-qubit gate fidelities of $0.993(1)$, and two-qubit gate fidelities of $0.9945(6)$ after correcting for losses during the gate operation.
        Furthermore, we present a novel state-resolved detection scheme for the two fine-structure states that enables high-fidelity detection of qubit loss.
        Finally, we leverage the existence of a stable ground state outside the qubit subspace to perform mid-circuit erasure conversion using fast destructive imaging.
        Our results establish the strontium fine-structure qubit as a promising candidate for near-term error-corrected quantum computers, offering unique scaling perspectives.
        \end{abstract}
	
    \maketitle
    
    Arrays of trapped neutral atoms are emerging as a leading platform for quantum computation~\cite{Saffman2010, Saffman2016, Henriet2020, Morgado2021}.
    Recent milestones include gate fidelities beyond $0.999$ for single-qubit gates~\cite{Sheng2018, Graham2022, Ma2023, Nikolov2023, Muniz2025, Yan2025} and around $0.995$~\cite{Evered2023, Ma2023, Radnaev2024, Finkelstein2024, Muniz2025} reaching up to $0.997$~\cite{Tsai2025} for two-qubit gates.
    Atom arrays have been scaled to thousands of qubits~\cite{Pause2024,Pichard2024,Manetsch2024,Tao2024,Gyger2024,Norcia2024,Lin2024}, and small-scale quantum computations, with both local~\cite{Graham2022} and nonlocal connectivity, have been demonstrated~\cite{Bluvstein2022}.
    This progress, together with recent demonstrations of logical circuits~\cite{Bluvstein2024,Reichardt2024,Bedalov2024}, has brought quantum error correction to the center of attention.

    A major obstacle to realizing logical qubits is the substantial overhead in the number of physical qubits required for their implementation.
    Noise-biased qubits have recently been proposed as a way to reduce this overhead~\cite{Tuckett2018,Tuckett2020,Lescanne2020, cong2022, Xu2023}.
    A particularly useful type of noise bias is towards leakage out of the computational subspace, which can be converted to a detectable error at a known location---thereby significantly increasing the threshold for quantum error correction~\cite{Sahay2023}.
    Such \textit{erasure-convertible qubits}, first demonstrated in atoms~\cite{Ma2023,Scholl2023}, have since been adopted by other platforms~\cite{Kubica2023,Alase2024, Chou2024, Holland2024, Levine2024, Quinn2024}.

    In atoms, erasure-convertible qubits are naturally implemented in metastable electronic states as found in alkaline-earth(-like) species.
    They were first proposed and realized in neutral ytterbium-171---where a nuclear spin-1/2 qubit is encoded in a metastable state~\cite{Wilson2022,Ma2023,Lis2023}---and have since been demonstrated in trapped ions~\cite{Kang2023,Quinn2024}.
    Implementing erasure-convertible qubits in other alkaline-earth atoms has proven more challenging.
    In strontium isotopes, the nuclear spin is either zero for bosonic strontium, or $9/2$ in the case of fermions, whose use in quantum computing applications based on qubits requires additional control demonstrated so far only for the electronic ground state~\cite{Barnes2022}.
    As a result, digital quantum computing with strontium atoms---for which high-fidelity imaging~\cite{Covey2019,Tao2024}, continuous atom reloading~\cite{Gyger2024}, and state-of-the-art gate fidelities~\cite{Finkelstein2024,Tsai2025} have been achieved---still mainly relies on the optical clock qubit~\cite{Finkelstein2024}.
    There, scaling to large system sizes and deep circuits may be challenging due to the limited gate speeds and high optical power requirements.
    Moreover, erasure conversion has so far only been applied to a short-lived qubit in strontium~\cite{Scholl2023} and not yet to any longer-lived encoding, in the context of quantum information processing.

    \begin{figure*}
    \centering
    \includegraphics[width=\textwidth]{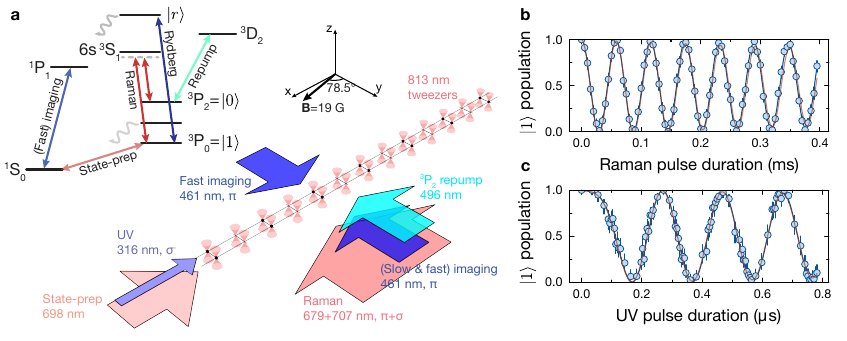}
    \caption{
    \textbf{Experimental setup.}
    \textbf{a.} Layout of the optical tweezer array, laser beams with their polarization, as well as the level diagram relevant to this work. 
    The qubit is encoded in metastable states $\ket 0 \equiv \ket{{}^3\text{P}_{2, m_J=0}}$ and $\ket 1 \equiv \ket{{}^3\text{P}_0}$ in ${}^{88}$Sr. 
    Repumping on the ${}^3\text{P}_2$-${}^3\text{D}_2$ transition at 496\,nm incoherently transfers $\ket 0$ population to ${}^1\text{S}_0$, in the state-sensitive readout scheme.
    Imaging in general is performed by scattering photons on the broad ${}^1\text{S}_0$-${}^1\text{P}_1$ transition at 461\,nm.
    For mid-circuit erasure detection of leakage out of the qubit subspace, two well-balanced, counter-propagating imaging beams are used.
    A tightly-focused UV beam couples $\ket 1$ to the Rydberg state $\ket r=\ket{47\text{s}\ {}^3\text{S}_{1, m_J=-1}}$ for entanglement generation.
    \textbf{b.} Rabi oscillation between $\ket 0$ and $\ket 1$ driven by a two-photon Raman process with (blue) and without (orange)  erasure conversion (discussed in the main text). 
    \textbf{c.} Rabi oscillation between $\ket 1$ and $\ket r$ with (blue) and without (orange) erasure conversion.
    Atoms in $\ket r$ are detected via losses induced by a $408\,$nm autoionization laser~\cite{Madjarov2020}.
    }
    \label{fig:1}
\end{figure*}

Here, we propose and experimentally demonstrate an erasure-convertible qubit in bosonic strontium-88, encoded in the fine-structure states ${}^3\text{P}_0$ and ${}^3\text{P}_2$ separated by $17\,$THz.
Compared to the qubit operating on the ultra-narrow clock transition~\cite{Finkelstein2024}, our scheme reduces the gate time by an order of magnitude while significantly lowering the required optical power, enabling straightforward scaling to larger array sizes.
Although coherent two‐photon coupling~\cite{Unnikrishnan2024,Pucher2024} and excellent qubit coherence~\cite{Ammenwerth2024} have been demonstrated for the fine-structure qubit, high-fidelity universal 
gates have remained elusive. 
We fill this gap by demonstrating and benchmarking a universal gate set on the fine-structure qubit.
First, we demonstrate coherent manipulation of fine‐structure qubits in an atom array via a two‐photon transition that couples the qubit states off‐resonantly through the intermediate ${}^3\mathrm{S}_1$ state. 
Next, we implement mid-circuit erasure conversion, converting state-preparation and off-resonant scattering errors into erasures and excising them to reach a single-qubit Clifford gate fidelity of $0.993(1)$.
Excluding loss events, we then realize two‐qubit gates based on Rydberg blockade with a fidelity of $0.9945(6)$, and generate Bell states of two fine‐structure qubits with a fidelity of $0.983(8)$. 
Finally, we describe in detail our loss detection scheme employed above, which leverages shelving into the ${}^1\text{S}_0$ state outside the qubit subspace, combined with controlled, low-leakage depumping, to enable imaging of both qubit states.
These results establish the fine‐structure qubit as a compelling candidate for neutral‐atom quantum computing and error correction.

Our experiment~\cite{Tao2024, Gyger2024, Ammenwerth2024} consists of strontium-88 atoms trapped in $32$ reconfigurable optical tweezers at $813\,$nm with spacings of $\SI{6.5}{\micro\meter}$ ($\SI{2}{\micro\meter}$) along the $x$ (or $y$) axis, see~\figref{fig:1}{a}. 
After loading atoms from the magneto-optical trap and performing initial imaging to determine tweezer occupancy, we sort them into a unity-filled pattern of pairs with a $\SI{13}{\micro\meter}$ inter-pair spacing using an additional mobile tweezer.
Subsequently, the atoms are cooled close to their radial motional ground state via resolved sideband cooling on the ${}^1\text{S}_0$-${}^3\text{P}_1$ transition~\cite{SI}.
The fine-structure qubit is encoded in metastable states $\ket {{}^3\text{P}_0} \equiv \ket 1$ and $\ket{{}^3\text{P}_{2, m_J=0}} \equiv \ket 0$. 
To prepare the state $\ket 1$, we apply a single, \SI{150}{\micro\second}-long $\pi$-pulse on the ultranarrow clock transition at $698\,$nm at $420\,$G bias field. 
To achieve magic trapping conditions for the fine-structure qubit, we then ramp the magnetic field to the triple magic configuration where the ${}^1\text{S}_0$ and the qubit states are magically trapped~\cite{Ammenwerth2024}.
For loading and imaging, we typically operate the tweezers at a depth of \SI{700}{\micro\kelvin}, which is reduced to a shallower depth of \SI{50}{\micro\kelvin} during gate operations to increase the qubit coherence time~\cite{Ammenwerth2024}.
To realize arbitrary single-qubit control, we drive a global two-photon Raman process using lasers at $679\,$nm and $707\,$nm, see~\figref{fig:1}{a,b}. 
To entangle pairs of atoms, a tightly focused $316\,$nm UV beam couples the state $\ket 1$ to a highly-excited Rydberg state $\ket r = \ket{47\text{s}\ {}^3\text{S}_{1, m_J=-1}}$ in free space (\figref{fig:1}{c}), where Rydberg interactions lead to a blockade of nearby excitations. 
In our array, the interaction shift for pairs of atoms in the Rydberg state is approximately $2\pi\times \SI{114}{\mega\hertz}$, which far exceeds the the Rabi coupling strength to the Rydberg state of $\Omega = 2\pi\times\SI{6}{\mega\hertz}$.
%

\begin{figure}[h!]
    \includegraphics[scale=1]{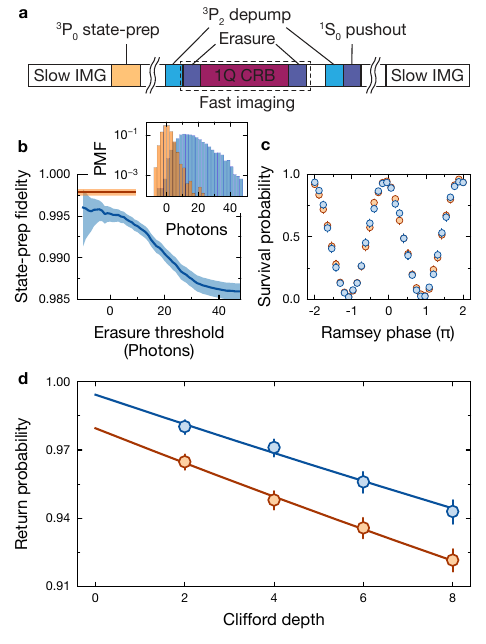}
    \caption{
    \textbf{State-preparation and single-qubit gate benchmarking.} 
    \textbf{a.} Experimental sequence. 
    The cooling-assisted, non-destructive imaging (slow IMG) at the start of the sequence verifies the loading of atoms at the target tweezer sites, and at the end reads out the $\ket 1$ population after a $\ket 0$ depletion stage (using ${}^3\text{P}_2$ depumping and ${}^1\text{S}_0$ pushout). 
    Two erasure conversion pulses sandwiching the single-qubit Clifford randomized benchmarking (CRB) sub-sequence remove state-preparation error in ${}^3\text{P}_0$ and off-resonant scattering errors originating from ${}^3\text{P}_1$. 
    See main text for details. 
    \textbf{b.} The $\ket 1$ state-preparation fidelity (blue) as a function of the erasure conversion threshold, which determines the fraction of state-preparation errors excised.
    Inset: Probability mass function (PMF) of photon counts in an erasure image, binarized using a preceding high-fidelity, low-loss imaging stage.
    %
    At low erasure threshold, the false negative probability is low, and the state-preparation error (blue) is suppressed to $0.39_{-0.21}^{+0.47}\%$, approaching the error from the imaging loss (red) alone.
    Negative photon counts arise solely from a camera bias that we subtract during post‐processing.
    \textbf{c.} Ramsey measurement with mid-circuit erasure conversion (blue) shows no measurable contrast reduction within the error bar compared to a reference measurement without it (orange). 
    \textbf{d.} The average Clifford gate fidelity raw (orange) and after erasure conversion (blue) from randomized benchmarking are $0.992(1)$ and $0.993(1)$, respectively.
    }  
    \label{fig:2}
\end{figure}

Since the ${}^1\text{S}_0$ ground state lies outside the qubit subspace, it can be used for erasure conversion via fast imaging on the broad ${}^1\text{S}_0$-${}^1\text{P}_1$ transition.
This has been demonstrated for the metastable, nuclear qubit in ytterbium-171~\cite{Ma2023}, a short-lived qubit encoded between the ${}^3\text{P}_0$ state and a Rydberg state in strontium-88~\cite{Scholl2023} and, more recently, the ground-state hyperfine qubit in strontium-87~\cite{Ma2025}. 
In our implementation, we image the ${}^1\text{S}_0$ population in a \SI{30}{\micro\second}-long detection block using two pulsed, counter-propagating imaging beams~\cite{Bergschneider2018, Su2025}, see~\figref{fig:1}{a}.
%
Based on the thresholding of the erasure image, we identify experimental instances where mid-circuit leakage to ${}^1\text{S}_0$ has occurred, and excise these from the analysis.
With a typical classification fidelity of $0.96$ for fast imaging in shallow tweezers during qubit operation, we can convert and excise approximately $91\,\%$ leakage error at the expense of discarding $7\,\%$ valid data where no error has occurred, see \figref{fig:2}{b} inset and \cite{SI}.
In practice, the erasure threshold can be adjusted during data processing to trade off between a lower error rate and a higher fraction of retained data. 
Here, we show that state-preparation errors arising from imperfect $\pi$-pulses on the clock transition and from Raman scattering during the idle time atoms spend in traps can be eliminated, see \figref{fig:2}{b}.
As we lower the classification threshold in the erasure image, the probability of successfully identifying and removing leakage errors increases.
As a direct consequence, the state-preparation fidelity increases. 
As the threshold drops below zero, the state-preparation fidelity reaches a plateau at over $0.996$, just $0.002$ below our imaging-limited survival probability of $0.998$. 
Moreover, we observe negligible influence of erasure conversion on the coherence of the fine-structure qubits, see~\figref{fig:2}{c}.
This establishes erasure conversion as a versatile, mid‐circuit‐applicable error‐detection scheme for the metastable qubits.

To demonstrate control over the fine-structure qubit initialized in $\ket 1$, we drive highly coherent Rabi oscillations between the two qubit states, see~\figref{fig:1}{b}.  
Our erasure scheme is also effective against leakage errors due to off-resonant scattering off ${}^3\text{S}_1$ into ${}^3\text{P}_1$, which quickly decays to the $^1\text{S}_0$ owing to its \SI{21}{\micro\second} lifetime. 
To perform state-sensitive detection of $\ket 1$, we drive the $496\,$nm repump transition between ${}^3\text{P}_2$ and ${}^3\text{D}_2$ to deplete $\ket 0$. 
The branching of ${}^3\text{D}_2$ into ${}^3\text{P}_J$ is such that $99.965\%$ of the population decays to $^1\text{S}_0$ via ${}^3\text{P}_1$, leaving $0.035\%$ in ${}^3\text{P}_0$, which amounts to a small detection error~\cite{Stellmer2014}.
Finally, we push out the population in $^1\text{S}_0$ with an intense $461\,$nm pulse and image the remaining population, which we assign to $\ket 1$.
%

To characterize the single-qubit gate fidelity, we turn to Clifford randomized benchmarking (CRB)~\cite{pygsti, Nielsen2020}. 
Specifically, we prepare atoms in $\ket 1$, apply a variable number of Clifford gates collectively equivalent to an identity operation in the absence of gate errors, before measuring the probability to find the qubit in state $\ket 1$ again, see~\figref{fig:2}{a}.
%
%
We find the average single-qubit Clifford gate fidelity $F_{\text{1q}}$ of 0.992(1) (0.993(1)) without (with) erasure excision, averaged over 16 tweezer sites, at $2\pi \times 17\,$ kHz Rabi frequency, see~\figref{fig:2}{d}.
Note that, in principle, the additional off-resonant scattering into ${}^3\text{P}_{2, m_J \neq 0}$ can be converted to erasure events by two-photon coupling of ${}^3\text{P}_2$ to the short-lived ${}^3\text{P}_1$ state.

\begin{figure}
    \includegraphics[scale=1]{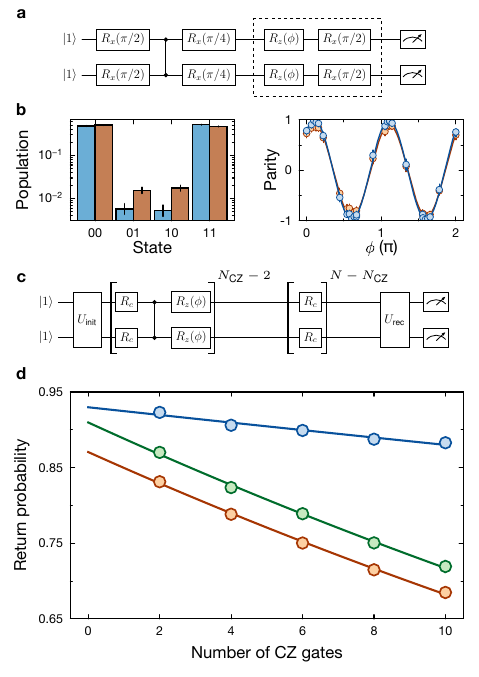}
    \caption{
    \textbf{Benchmarking 2-qubit gates.} 
    \textbf{a.} 
    Circuit for generating a maximally entangled Bell state $\ket \Phi^+ = (\ket{00} + \ket{11})/\sqrt{2}$ and probing its fidelity. 
    $R_z(\phi)$ is implemented virtually via frame tracking.
    The dashed box denotes the analyzer circuit that induces parity oscillation.
    \textbf{b.} Bell state tomography and parity oscillation together yield a Bell state fidelity of $0.983(8)$ (blue) and $0.935(9)$ (red) with and without post-selection on the survival of the atom pairs. 
    \textbf{c.} SSB sequence used to benchmark the CZ gate fidelity. 
    $R_c$ denotes $\pi/2$-pulses with random phases. 
    $U_{\text{init}}$ and $U_{\text{rec}}$ denote initialization and recovery gate sequence, respectively~\cite{Finkelstein2024}.
    \textbf{d.} The CZ gate fidelity from fitting decaying $\ket{11}$ return probability in a SSB sequence. Here, the raw (red), erasure-converted (green), and loss-excised (blue) fidelity are 0.9759(5), 0.9764(8), and 0.9945(6), respectively.
    }
    \label{fig:3}
\end{figure}

An important requirement for universal quantum computers is the realization of high-fidelity entangling gates.
To entangle two fine-structure qubits, we couple the qubit state $\ket{1}$ to the Rydberg state $\ket{r}$, implementing a time-optimal control-Z (CZ) gate by modulating the phase of the laser drive~\cite{Jandura2022}. 
Embedding the CZ gates in a sequence of single-qubit rotations (see~\figref{fig:3}{a}), we benchmark the entanglement by generating two-qubit Bell states.
From parity oscillation contrast and population measurements in the $\ket{00}$ and $\ket{11}$ subspace, we extract a Bell-state fidelity of $0.9834$ ($0.9355$) with (without) excising atom loss events using the procedure described below, see~\figref{fig:3}{b}.
To obtain a CZ-gate fidelity $F_{\text{2q}}$, we implement the recently proposed symmetric stabilizer benchmarking (SSB) sequence~\cite{Finkelstein2024, Tsai2025}, which is only weakly sensitive to single-qubit gate errors.
By fitting the decaying $\ket{11}$ return probability as a function of number of CZ gates as in $P_{11}=a F_{\text{2q}}^{N_{\text{CZ}}}$, we measure $F_{\text{2q}}=$0.9945(6) (0.9759(5)) with (without) correction for atom loss events, see~\figref{fig:3}{c,d}. 

\begin{figure}
    \includegraphics[scale=1]{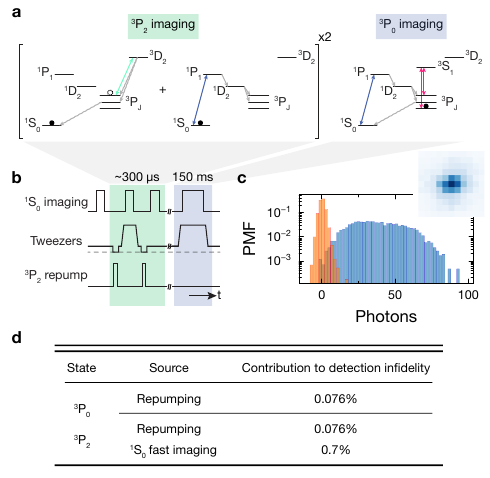}
    \caption{
    \textbf{State-resolved detection}.
    \textbf{a.}
    Internal-state dynamics relevant to the detection of ${}^3\text{P}_2$ (green) and ${}^3\text{P}_0$ (blue). 
    Prior to detecting ${}^3\text{P}_2$, population in this state is incoherently transferred to $^1\text{S}_0$ via repumping on the ${}^3\text{P}_2$-${}^3\text{D}_2$ transition.
    Then, we perform fast, destructive imaging in deep tweezers to collect sufficient photons for high‐fidelity ${}^3\text{P}_2$-state reconstruction.
    Such a state-transfer plus imaging cycle is repeated to ensure complete removal of ${}^3\text{P}_2$ population, despite small leakage from ${}^1\text{P}_1$ via ${}^1\text{D}_2$ into ${}^3\text{P}_2$.
    Once ${}^3\text{P}_2$ is destructively imaged, the remaining population is assigned to ${}^3\text{P}_0$ using a subsequent slow image with repumping on the $679\,$nm and $707\,$nm transitions.
    \textbf{b.} Timing diagram (not to scale) for the ${}^1\text{S}_0$ imaging beam, ${}^3\text{P}_2$ repumper, and tweezer trap depth during the state-selective detection.
    Traps are blinked off (dashed) during the ${}^3\text{P}_2$ repumping pulse to avoid photo-induced losses.
    \textbf{c.} Probability mass function (PMF) of photons emitted by $^1\text{S}_0$ atoms transferred from ${}^3\text{P}_2$ recorded during fast imaging.
    The resulting histogram, binarized using a high-fidelity pre-image, shows a classification fidelity $>0.993$ averaged over all tweezers at the optimal threshold. 
    The orange and blue correspond to zero- and one-atom peak, respectively.
    Inset: deconvolution mask derived from the averaged fluorescence profile.
    \textbf{d.}
    Dominant infidelity contributions to state-resolved imaging.
    }
    \label{fig:4}
\end{figure}

To decouple the entangling gate fidelity from qubit loss (i.e.\ due to ionization of $^3\text{P}_2$ by the 316 nm Rydberg excitation laser \cite{SI}), we present a new, partially destructive, state-resolved detection (SRD) scheme for the fine-structure qubit.
Contrary to the previous work focusing on imaging one of the two states in a qubit~\cite{Fuhrmanek2011, Gibbons2011, Martinez2017, Kwon2017}, the SRD yields population in both qubit states and can thus identify atom loss events.
In alkali species, direct imaging of both qubit states has been achieved by spatially separating spin states in strong magnetic field gradients~\cite{Boll2016, Koepsell2020} or state-dependent traps~\cite{Wu2019}. 
In alkaline-earth(-like) atoms, the SRD was recently demonstrated by imaging on a narrow-line transition and shelving in metastable clock states~\cite{Norcia2023, Lis2023}. 
To convert the state-sensitive detection scheme including the $^3\text{P}_2$-repumper into the SRD, we collect the photons emitted during the push of $^1\text{S}_0$ using fast imaging techniques following the ${}^3\text{P}_2$-${}^3\text{D}_2$ repumping step, see~\figref{fig:4}{a,b}.
%
Compared to the fast imaging used for erasure conversion, the photon count here is increased by operating at a significantly higher tweezer trap depth.
%
We achieve $>0.9931$ detection fidelity of ${}^1\text{S}_0$ atoms in $\sim$\SI{50}{\micro\second}, which significantly exceeds the fidelity achieved in shallow traps for state-preparation erasure conversion, see~\figref{fig:4}{c}.
Meanwhile, fast imaging removes $>99.92\%$ of the population from $\ket 0$, and ensures high-fidelity assignment of the remaining atoms to $\ket 1$ in a subsequent conventional imaging step. 
As a result, we find a $\ket 0$ ($\ket 1$) detection fidelity $>0.993$ ($0.998$), see~\figref{fig:4}{d}.
Note that our scheme is minimally sensitive to bit-flip errors originating from trap-induced Raman processes. 
This is a consequence of the short timescale and relatively shallow traps during which the SRD operates, in contrast to alternative cooling-dependent schemes~\cite{Norcia2023}.
%

In summary, we have demonstrated universal quantum control of the 17\,THz fine-structure qubit in strontium-88. 
In addition to the demonstrated erasure conversion, this qubit has several distinct advantages over existing approaches.
First, by employing larger intermediate-state detunings and higher optical powers, single-qubit gate times could be driven well below those demonstrated in our proof-of-principle experiment—potentially into the sub-\SI{}{\micro\second} regime in future implementations.
Compared to the hyperfine or nuclear qubits, in particular the metastable nuclear qubit in ytterbium-171, the fine-structure qubit permits larger intermediate-state detunings and thus in principle proportionally smaller error per gate.
Second, we expect that the fine-structure qubit is compatible with coherent transport~\cite{Bluvstein2022}, paving the way for universal quantum computation and error correction~\cite{Bluvstein2024} enhanced by erasure conversion~\cite{Wu2022}. 
Third, we expect that our state-resolved detection will facilitate monitoring inevitable atom loss in quantum computation---for example, during error-correction cycles~\cite{Chow2024,Yu2024,Perrin2024}.
To this end, the residual infidelity in our state-resolved detection scheme could be mitigated by adding repumping lasers on the $^1\text{D}_2$ or by deploying a more efficient imaging system.
Finally, we anticipate that combining our state-resolved detection scheme with fast, MHz-scale fine-structure qubit manipulation will unlock a broader range of applications.
For example, this combination enables readout of population in short-lived transitions to $^3\text{P}_2$, such as the decay of $^3\text{D}_3$ to $^3\text{P}_2$ on the long-wavelength \SI{2.9}{\micro\meter} transition relevant for free-space quantum optics experiments~\cite{Masson2024}.
Another application is the detection of both qubit states of short-lived qubits in Rydberg-based quantum simulators, enabling high-fidelity quantum simulation that is robust to atom loss~\cite{Scholl2021, Ebadi2021, Scholl2023}. 

\begin{acknowledgments}
		We thank Giuliano Giudici, Giacomo Giudice for support in theoretically modeling and optimizing CZ gates, Simon Evered for experimental insight in gate parameter optimization, Rick van Bijnen, Andreas Kruckenhauser for the calculation of Rydberg interaction potentials, David Wei, Taylor Briggs, Niklas Zischka for proofreading the manuscript, and Elias Trapp for early assistance in building the \SI{813}{\nano\meter} laser system.
		We acknowledge funding by the Max Planck Society (MPG) the Deutsche Forschungsgemeinschaft (DFG, German Research Foundation) under Germany's Excellence Strategy--EXC-2111--390814868, from the Munich Quantum Valley initiative as part of the High-Tech Agenda Plus of the Bavarian State Government, and from the BMBF through the programs MUNIQC-Atoms and MAQCS.
		This publication has also received funding under Horizon Europe programme HORIZON-CL4-2022-QUANTUM-02-SGA via the project 101113690 (PASQuanS2.1).
		J.Z. acknowledges support from the BMBF through the program “Quantum technologies---from basic research to market” (SNAQC, Grant No. 13N16265).
		H.T., M.A. and R.T. acknowledge funding from the International Max Planck Research School (IMPRS) for Quantum Science and Technology. M.A acknowledges support through a fellowship from the Hector Fellow Academy.
		F.G. acknowledges funding from the Swiss National Fonds (Fund Nr. P500PT\underline{\hspace{2mm}}203162).
        O.L. acknowledges support from the Rothschild and CHE Quantum Science and Technology fellowships.
		
	\end{acknowledgments}

\appendix

\setcounter{figure}{0}
\renewcommand\theequation{S\arabic{equation}}
 \renewcommand\thefigure{S\arabic{figure}}

\section{Supplementary Information}
\section{Details about the experimental apparatus}
    Our experiment including MOT loading, motional ground-state cooling, and imaging has already been described in prior work~\cite{Tao2024, Gyger2024, Ammenwerth2024}.
    Here, we only present new developments connected to this work.
    
    \subsection{Tweezers equalization and imaging} 
\begin{figure*}[t]
    \centering
    \includegraphics[scale=1]{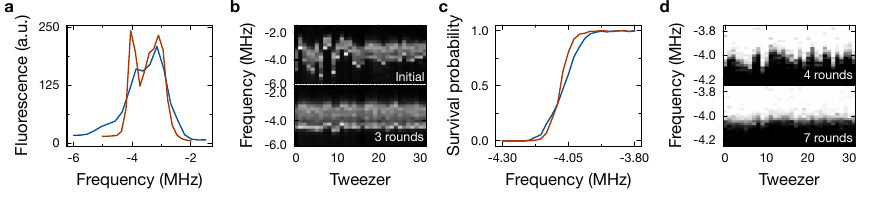}
    \caption{
    \textbf{Tweezer trap depth equalization baesd on Sisyphus cooling on ${}^1\text{S}_0$-${}^3\text{P}_1$ transition.} 
    \textbf{a.} Tweezer-averaged fluorescence shows pronounced spectral narrowing from uncorrected tweezers (blue) in tweezers corrected with 3 rounds of amplitude feedback (red).
    \textbf{b.} When tweezer-resolved features are well separated initially, the mean cooling frequency that maximizes the fluorescence is used to balance the amplitude difference between tweezers. 
    The numbers label the index of the iteration round.
    \textbf{c.} After a few iterations of fluorescence-based feedback, tweezers become sufficiently homogeneous and reasonably high imaging fidelity can be reached. 
    For further optimization, we focus on the blue-detuned edge of the cooling feature in an imaging block owing to its better correlation to trap depth in the attractive cooling regime.
    Arrays with better amplitude homogeneity (red) exhibit sharper transition between low and high survival in imaging than the ones with lower homogeneity (blue).
    \textbf{d.}
    For feedback purpose, the midpoint of the slope between low and high survival probability is used. 
    The numbers label the index of the iteration round, continuing from the labeling in the fluorescence-baesd optimization.
   }
    \label{fig:tw_equalization}
    \end{figure*}

    For equalizing tweezer trap depths, we leverage an observed correlation between the optimal Sisyphus cooling condition on the $^1\text{S}_0$ to $^3\text{P}_1$ transition and tweezer trap depth. 
    This probe remains robust even in the early stages of optimization, where the tweezer array exhibits large variations in trap depth that renders imaging on the global scale challenging.
    We begin with a tweezer array that is sufficiently uniform in trap depth to enable atom loading in every site.
    If this condition is not met, an initial optimization using a separate low-NA imaging system with a camera is typically necessary.
    We then scan the Sisyphus cooling frequency during imaging and identify the frequency that maximizes fluorescence for each tweezer.
    This site-resolved optimal cooling frequency serves as an experimental observable that can be used for equalization, see~\figref{fig:tw_equalization}{a,b}.

\begin{figure*}[t]
    \centering
    \includegraphics[scale=1]{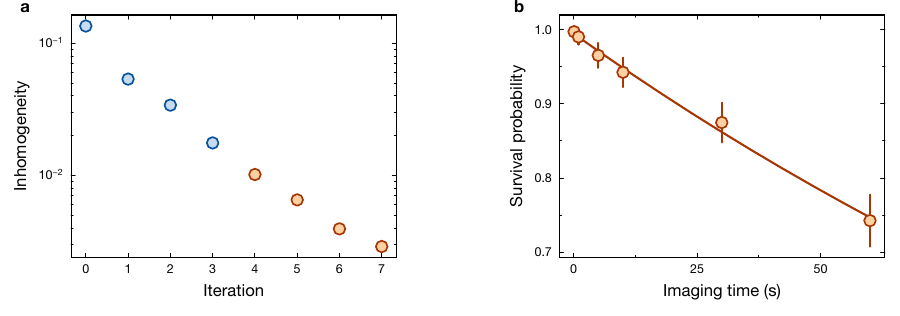}
    \caption{
    \textbf{Convergence plot for tweezer amplitude equalization and imaging characterization} 
    \textbf{a.}
    The atomic signal used is initially the central cooling frequency, which maximizes fluorescence during iterations 0–3, and later the blue‐detuned edge of the cooling frequency, which optimizes survival probability during iterations 4–7.
   \textbf{b.}
   With a tweezer array optimized by this homogenization strategy, the continuous imaging lifetime exceeds 200 seconds, corresponding to a survival probability of $>0.99928$ for a 150\,ms-long imaging.
   }
    \label{fig:convergence}
    \end{figure*}

    To equalize trap depth below the percent level, we try to overlap the steep edge of the Sisyphus cooling feature at blue detunings, and sharpen the transition in survival probability from low to high, see~\figref{fig:tw_equalization}{c,d}.
    In holographic arrays created using a spatial light modulator (as in our case), this is achieved by tuning the weight of each tweezer in the phase-retrieval algorithm.
    Our protocol shows a strong convergence and is expected to further reduce inhomogeneity beyond $\sim0.3\%$ achieved here with more iterations, see~\figref{fig:convergence}{a}.
    To confirm a trap depth homogeneity across the entire array after the optimization, we show we can reach more than 209\,seconds lifetime under continuous illumination, see~\figref{fig:convergence}{b}. 
    As a comparison, our vacuum lifetime is estimated to be 273 seconds~\cite{Tao2024}.
    The imaging lifetime corresponds to an imaging survival probability $>0.99928$ for 150\,ms-long imaging.
    Note that the imaging survival probability is slightly lower for the measurement involving erasure conversion of state-preparation errors, due to using a different array with greater inhomogeneity.

    \subsection{Atom rearrangement}
    To rearrange atoms in a tweezer array generated by a spatial-light modulator (SLM), we use another, mobile tweezer generated by a two-axis, acousto-optical deflector (AOD). 
    The AOD and SLM beams are merged on a polarizing beam splitter, and sampled by a pick-up plate before a high-resolution objective focuses them into tweezers. 
    The two systems share a long optical path that plays an important role in reducing slow relative drift between moving and static tweezers. 
    The sampled tweezer light is imaged onto a CCD camera for daily adjustment of their relative alignment for optimal sorting performance.

    Our tweezer alignment calibration takes advantage of the common coordinate reference defined by the monitoring camera. 
    To calibrate the positions of the two arrays, we first apply a multi-tone radio-frequency (RF) drive to the AOD to create a 3-by-3 tweezer array. 
    Then, we record the tweezers' positions and compute the affine transformation $M$ that converts an AOD tweezer's coordinate into the required RF tone. 
    %
    $M$ accounts for rotation, scaling, shearing, and translation that typically occur in the transformation between the AOD’s spectral domain and the tweezers’ spatial domain in the camera image.
    Finally, we display and image the SLM tweezer array to get the absolute coordinate of static tweezers. 
    This step completes the calibration and allows us to accurately overlap an AOD tweezer with any SLM tweezer, for as long as a few days without the need to recalibrate. 
    We find that the result from tweezer calibration on the monitor camera directly carries over to the atomic plane, such that any additional atomic-based tweezer overlapping is unnecessary. 

    To rearrange the atoms into a target configuration, we take a high-fidelity fluorescence image of a stochastically loaded SLM array to obtain the initial density distribution. 
    Then, we run a heuristic rearrangement algorithm to calculate a list of trajectories that reshuffle atoms into a defect-free target array. 
    In three straight moves, our algorithm picks up, transports, and deposits the atoms at the target site.
    For small arrays (e.g.\ 2-by-4), the probability of assembling a defect-free array is $0.955$.  
    We attribute such a high-fidelity sorting to high-fidelity imaging which has survival probability exceeding $0.999$ and the practice of maintaining a constant RF load on the AOD during the idle time to reduce thermally induced beam-pointing instability.

    \begin{figure*}[t]
    \centering
    \includegraphics[scale=1]{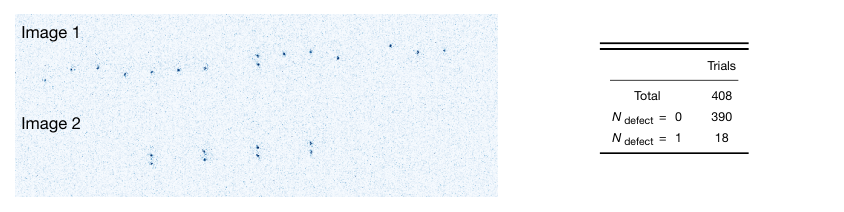}
    \caption{
    \textbf{Atom rearrangement}
    Example atomic occupation in fluorescence images before (image 1) and after (image 2) atom rearrangement. 
    In image 1, atoms are loaded into 32 tweezers stochastically.
    The sorting procotol fills the target tweezer sites (2-by-4 array) in image 2, and remove any atom surplus. 
    The probability of assembling defect-free 2-by-4 arrays is $0.955$. 
   }
    \label{fig:sorting}
\end{figure*}

    \subsection{Ground-state cooling and state preparation}
\begin{figure*}[t]
        \centering
        \includegraphics[scale=1]{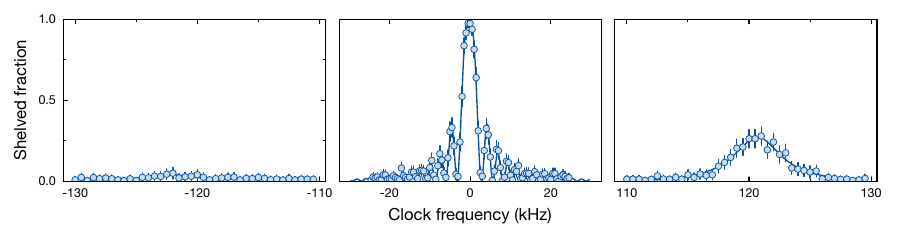}
        \caption{
        \textbf{Sideband thermometry on the narrow clock transition.}
        The probe time $t_\text{probe}$ = $3t_\pi$ where $t_\pi$ is the $\pi$-time on the carrier transition. The mean motional occupation $\bar n$ averaged over 32 tweezers is $<0.1$, which amounts of $>90\,\%$ probability of finding atoms in absolute motional ground state.
       }
        \label{fig:sc}
    \end{figure*}

    To initialize the atom for the subsequent manipulations, we perform resolved sideband cooling on the $689\,$nm transition in SLM tweezers. 
    By tuning the polarizability using a magnetic bias field, the differential light shift of the cooling transition vanishes, and the cooling frequency becomes independent of the atom’s position in the trap~\cite{Norcia2018,Norcia2019}. 
    There are two sideband cooling beams installed in our apparatus. 
    The radial sideband cooling beam has $\vec{k}$ largely within the atomic plane of the tweezers, and effectively couples only to the radial sidebands. 
    Another sideband cooling beam propagates through the high-resolution objective and couples axial sidebands. 
    We achieve largely uniform illumination on the atomic plane by weakly focusing the axial cooling beam onto the back-focal plane of the objective. 
    To optimize cooling, we scan various parameters to maximize the survived fraction of the cooled atoms after a fixed release-and-recapture time. Better cooling leads to higher recapture rate. 
    Our cooling sequence consists of 5 ms of radial sideband cooling only, and then alternating radial + axial sideband cooling for another 25 ms. 
    To measure the effective temperature of the atoms after cooling, we perform sideband thermometry on the narrow clock transition, and consistently obtain $\bar n < 0.1$ along the probed direction, see~\figref{fig:sc}{}.

\begin{figure*}[t]
        \centering
        \includegraphics[scale=1]{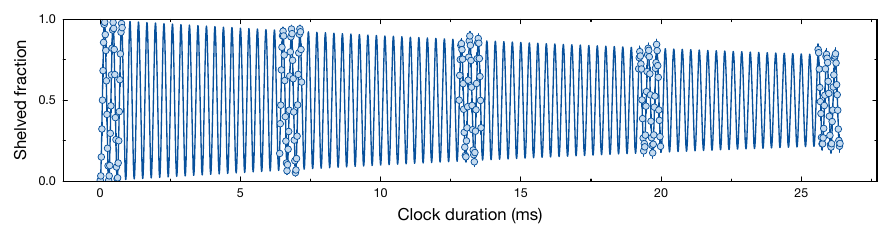}
        \caption{
        \textbf{Rabi oscillation on the clock transition.}
        A simple fit yields a Rabi frequency of $2\pi\times 3.3\,$kHz and dephasing of $0.022$/ms.
        The measurement is carried out in 420\,G magnetic field.
        }
        \label{fig:rabi}
    \end{figure*}

    We initiate atoms in $\ket 1$ via a simple, resonant $\pi$-pulse on the ultra-narrow clock transition.
    We benchmark the expected state-preparation fidelity by measuring a Rabi drive on the clock.
    With up to 420 G magnetic field and 10~mW optical power, we can reach $2\pi \times 3\,$kHz Rabi frequency, and 150 coherent cycles within $1/e$ dephasing time, see~\figref{fig:rabi}{}. 
    Our clock laser system consists of an ECDL, an injection-locked amplifier, and a high-finesse cavity with finesse $>200,000$. 
    To compensate the slow frequency drift due to the natural aging of ULE glass to which the seed laser is locked, we compare the clock frequency to a frequency comb referenced to a maser. 
    The instantaneous frequency correction is then applied to a modulator in the optical path of the clock laser after filtering to constantly maintain the resonance condition on the atoms. 
    We do not actively cancel fiber noise, despite using a 10 m fiber for light delivery to the experiment. 
    Typically, the state-preparation infidelity is less than $1\,\%$, and we further reduce it using state-preparation erasure.

    \subsection{Fast imaging and erasure conversion of state-preparation errors}
    To rapidly image atoms for erasure conversion, we implement the protocol first demonstrated in \cite{Bergschneider2018}  and expanded in \cite{Su2025}. 
    We illuminate atoms with a pair of intense, counter-propagating, $461\,$nm imaging beams. 
    The laser system is based on an in-house developed, injection-locked amplifier seeded by a Toptica SHG laser at the fundamental wavelength of 922\,nm. 
    We constantly monitor the state of the injection lock via observing the transmission of a scanning Fabry-Perot cavity and relock the system if the signal falls below a certain threshold. 

    On the experiment side, we have about 10 mW light in each of the counter-propagating arms of the imaging system.
    The polarizations of both imaging beams are largely parallel to the quantization axis defined by $19\,$G magnetic field to drive $\pi$-transition on the ${}^1\text{S}_0$-${}^1\text{P}_1$ line. 
    With a high-resolution objective ($\text{NA}=0.65$) conveniently placed $90^\circ$ relative to the quantization axis, we achieve an optimal photon collection efficiency of $0.16$ considering dipole emission pattern alone. 
    The imaging beams are resonant with the target transition in a trap with a depth of \SI{55}{\micro\kelvin}, to produce maximal scattering.
    We pulse the RF to the modulators so that the intensities $I_i, i=1, 2$ of both imaging beam are on in an alternating fashion to avoid uncontrolled spreading of momentum at saturation. 
    %

    \begin{figure*}[t]
        \centering
        \includegraphics[scale=1]{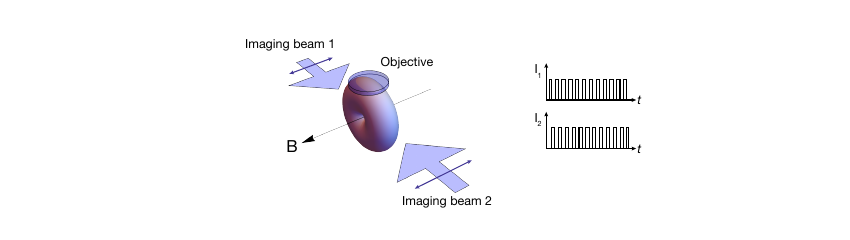}
        \caption{
        \textbf{Dipole emission pattern for efficient photon collection.} 
        The imaging beams are both linearly polarized (double-arrowed) relative to the quantization axis defined by the $19\,$G magnetic field (single-arrowed).
        The photon collection efficiency is about 0.16 when both imaging beams drive $\pi$-transition. 
        The intensities are pulsed in an alternating fashion to balance momentum kick. 
        }
        \label{fig:dipole}
    \end{figure*}

    \begin{figure*}[t]
        \centering
        \includegraphics[scale=1]{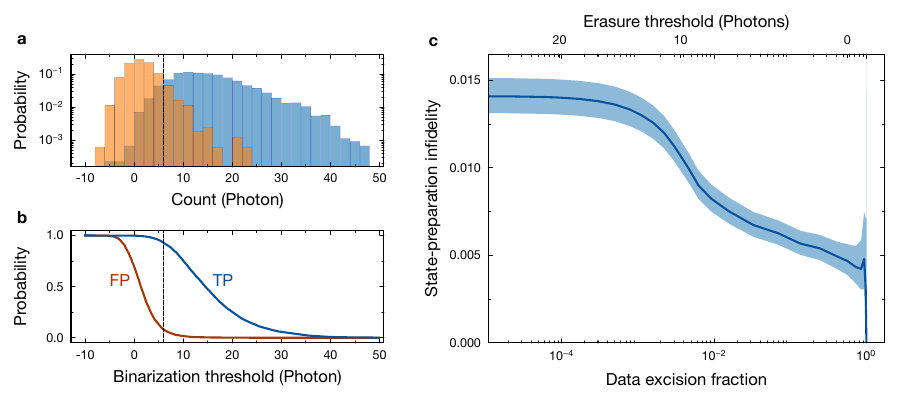}
        \caption{
        \textbf{Fast imaging classifier validation based on labeled data.}
        \textbf{a.} 
        The histogram of photon counts from an erasure‐detection image, binarized using a high‐fidelity, low‐loss pre‐image. 
        The orange bars correspond to counts when no atom was present in the pre‐image, and the blue bars correspond to counts when an atom was present. 
        We make a binary classification of whether an atom is detected in an erasure image by comparing the analog fluorescence counts against a threshold (dashed).
        \textbf{b.}
        In choosing an appropriate threshold, our primary figure of merit is the true positive (TP, blue) probability, which measures how often we correctly identify events with leakage errors. 
        Lowering the threshold captures a larger fraction of true positives but also increases the false positive (FP, red) probability, which is how often we misclassify events without leakage as if they had leakage.
        %
        %
        An usually optimal threshold maximizes the classification fidelity (dashed).
        \textbf{c.}
        In converting state-preparation error to erasure errors, we see the balancing between lower false positive and higher true positive probability by computing the state-preparation infidelity at various data excision fractions.
        At a low erasure threshold (i.e.\ 20 photons), most erasure images appear dark, indicating no leakage error in state-preparation.
        As a result, few experimental shots are excised and the state‐preparation infidelity remains the same as when no erasure conversion is applied. 
        Conversely, at a high erasure threshold (i.e.\ 5 photons), most leakage events produce bright erasure images.
        Consequently, removing all shots flagged as erasures lowers the apparent state‐preparation infidelity.
        }
        \label{fig:erasure_conversion_thresholding}
    \end{figure*}
    With fast imaging, we can detect ${}^1\text{S}_0$ population in approximately \SI{30}{\micro\second}, a prerequisite for mid-circuit erasure conversion.
    To benchmark the fidelity of atom detection in fast imaging, we first label each atom’s state using our slow imaging, and then pass the same atom to the fast‐imaging stage to validate the labeled data.
    This validation procedure is justified by the high classification fidelity and survival probability of our slow imaging.
    The fluorescence counts acquired in the fast imaging block usually follow a photon‐count distribution that differs depending on whether an atom is present or absent (see~\figref{fig:erasure_conversion_thresholding}{a}).
    One can then maximize classification fidelity by choosing a threshold that minimizes false positive and false negative probabilities, see~\figref{fig:erasure_conversion_thresholding}{b}. 
    In practice, the threshold is often biased toward a higher true positive probability for improved error conversion, at the expense of a proportionally higher false positive probability. 
    In the case of erasure conversion and excision of state-preparation error, this trade-off is easily seen, see~\figref{fig:erasure_conversion_thresholding}{c}.
    That is, to convert a larger fraction of leakage errors to erasure errors that can be excised for higher state-preparation fidelity, one has to discard more valid data in which no leakage error has occurred.
    This leads to a lower fraction of retained data.
    However, such a trade‐off is less severe in fast imaging with higher classification fidelity.
        
    \begin{figure*}[t]
        \centering
        \includegraphics[scale=1]{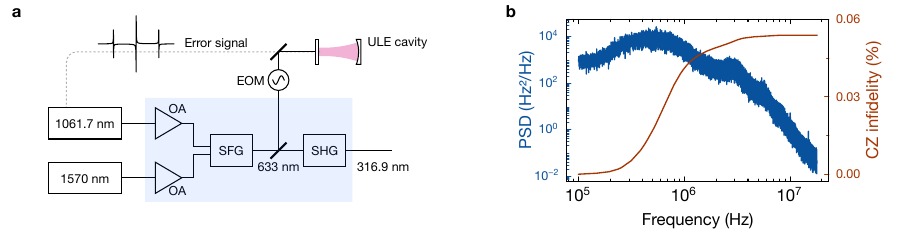}
        \caption{
        \textbf{UV laser system and frequency noise.} 
        \textbf{a.} The UV laser system including the PDH lock for frequency regulation. 
        \textbf{b.} The frequency noise power spectral density measured using the 633\,nm light and converted to that in UV. 
        The cumulated CZ infidelity (red) for $2\pi\times 6\,$MHz Rabi is estimated from frequency noise using linear repsonse theory~\cite{Tsai2025}. 
       }
        \label{fig:laser}
    \end{figure*}

    \subsection{$496\,$nm laser system}
    Our $496\,$nm laser for ${}^3\text{P}_2$ discrimination is a Moglabs ECDL with typically 10-20\,mW output power. 
    We lock the laser to a Moglabs Fizeau wavemeter to counteract slow frequency drift. 
    With usually a few mW optical power on atoms, we can deplete ${}^3\text{P}_2$ with up to $99.5\,\%$ fidelity in less than \SI{3}{\micro\second}. 
    The depumping error is mostly due to atoms shelved into the meta-stable states via branching off ${}^1\text{P}_1$ and can be significantly reduced with one more round of repumping and pushout pulses.
    While simple and effective, our spin-selective detection cannot discriminate between population in different Zeeman sublevels in ${}^3\text{P}_2$ due to strong saturation on the repuming transition. 
    As a result, the apparent $\ket 0$ population will be higher than it actually is due to the collateral removal of $m_J \neq 0$, the non-qubit states in a ${}^3\text{P}_2$ push-out.
    The population in the non-qubit states originates for example from scattering from ${}^3\text{S}_1$ in single-qubit gates or decay of the Rydberg state during the two-qubit gates.
    
    \subsection{UV laser system}
    To generate the ultraviolet (UV) light at $316\,$nm, we use two nonlinear mixing stages. 
    We start with two seeds at $1061.7\,$nm (extended-cavity diode laser, ECDL) and $1570\,$nm (fiber laser), which are then amplified in fiber amplifiers before converting into $~\sim4\,$W, $633\,$nm light via sum-frequency generation (SFG). 
    Subsequently, we convert $633\,$nm light into $316.9\,$nm via second-harmonic generation (SHG), with a peak UV power of $\sim800\,$ mW. 
    The laser module is under a closed-loop circulation of clean, dry air to maintain a contaminant-free environment critical to the laser's lifetime.  
    For pulse engineering, we employ two acousto-optic modulators (AOMs).
    The first AOM is a single-pass AOM regulating the DC intensity of the UV, while the second one modulates the pulse in a double-pass configuration.
    The beam at the second AOM is imaged to the atoms to faithfully relay the phase modulation to the atomic plane.

    To stabilize the UV frequency, we lock the $633\,$nm SFG system to a ULE cavity with finesse $52000$.
    The control signal derived from a Pound-Drever-Hall lock is applied to the $1061.7\,$nm ECDL. 
    We measure the frequency noise power spectral density $\text{PSD}_\text{red} [\text{Hz}^2/\text{Hz}]$ at 633\,nm using a delayed Mach-Zehnder interferometer~\cite{Denecker2024}. 
    We convert it to the frequency noise in the UV via $\text{PSD}_\text{UV} [\text{Hz}^2/\text{Hz}] = 4 \cdot \text{PSD}_\text{red} \cdot H_\text{SHG}^2$ accounting for the noise amplification in the second-harmonic generation and the SHG cavity transfer function $H_\text{SHG}$, see~\figref{fig:laser}{b}.

    We target the $n=47\,{}^3\text{S}_1, m_J = -1$ Rydberg state with our UV laser system. 
    The specific principal quantum number is chosen primarily for its reduced DC Stark shift (resonance frequency variation induced by external electric field) compared to higher $n$. 
    To further reduce shot-to-shot changes of stray charges in the vicinity of atoms, we flash our glass cell with an intense UV flash during each sequence. 
    To remove residual drifts, we implement an atomic servo which measures slow resonance drift and correct it in the science sequence.
    The in-loop error signal indicates a detuning error of as much as $400\,$kHz per day. 

    \section{Rydberg lifetime and branching ratio}
\begin{figure*}[t]
        \centering
        \includegraphics[scale=1]{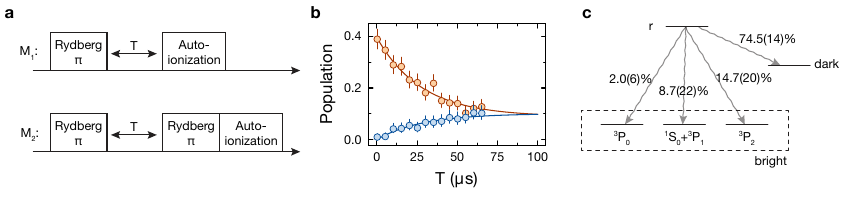}
        \caption{
        \textbf{Rydberg lifetime and branching into lower-lying states.}
        \textbf{a.}
        The two complementary sequences, when analyzed together, determine the lifetimes of Rydberg states decaying into both bright and dark channels.
        \textbf{b.}
        The population measured in sequence $M_1$ (blue) and $M_2$ (orange). 
        The solid lines are fit functions derived from the decay model, see main text.
        The dark and bright state lifetimes are \SI{37(2)}{\micro\second} and \SI{110(8)}{\micro\second}, respectively.
        The experiments are performed in free-space, resulting in the drop in population independent of $T$.
        \textbf{c.} Measured branching ratio into dark and low-lying states.
       }
        \label{fig:lifetime}
    \end{figure*}

    We measure the Rydberg state lifetime using the pair of sequences in \figref{fig:lifetime}{a}. 
    Inspired by recent work~\cite{Scholl2023, Cao2024}, we make a distinction between decays into bright states and dark states.
    Bright states---namely ${}^1\text{S}_0$ and the ${}^3\text{P}_J$ manifold---produce a positive signal in our standard fluorescence-based imaging.
    In contrast, the dark states produce no positive signal and appear dark. 
    For example, population in nearby Rydberg levels can be lost through anti‐trapping and thus cannot be detected.
    
    The population $P_1, P_2$ measured in \figref{fig:lifetime}{a} do not directly yield the lifetime of interest. 
    To establish the missing link, we employ a simplified description using a three-level system containing Rydberg state $\ket r$, dark state $\ket d$, and bright state $\ket b$. The dynamics during the decay can be described by equations 
    \begin{align*}
        P_r'(t) &= - \left(\frac{1}{\tau_b} + \frac{1}{\tau_d}  \right) P_r(t),\\ 
        P_d'(t) &= \frac{1}{\tau_d} P_r(t),\\
        P_b'(t) &= \frac{1}{\tau_b} P_r(t).
    \end{align*} 
    To streamline the analysis, we combine the dark-to-bright decay with the Rydberg-to-bright decay. 
    With initial condition $P_r(0) = A, P_d(0) = P_b(0) = 0$, the solution of the rate equations is
    \begin{align*}
        P_r(t) 
        &= A \exp\left[ -t \left( \frac{1}{\tau_d} + \frac{1}{\tau_b} \right)\right] \\ 
        P_d(t) 
        &= A \frac{\tau_b}{\tau_b + \tau_d} 
        \left(1 - \exp\left[ -t \left(  \frac{1}{\tau_d} + \frac{1}{\tau_b} \right) \right] \right) \\ 
        P_b(t) 
        &= A \frac{\tau_d}{\tau_b + \tau_d} 
        \left(1 - \exp\left[ -t \left(  \frac{1}{\tau_d} + \frac{1}{\tau_b} \right) \right] \right)
    \end{align*} 
    Here, $A$ is the global normalization factor which is about $0.4$ due to loss of atoms in a \SI{70}{\micro\second} long free-flight independent of $T$.

    It follows that 
    \begin{align}
        P_1(t) &= P_b(t) \\ 
        P_2(t) &= A - P_d(t) - \frac{1}{9} P_b(t).
    \end{align}
    The $1/9$ accounts for a re-excitation of ${}^3\text{P}_0$ population to $\ket r$ at the second Rydberg $\pi$-pulse and its subsequent removal by the auto-ionization pulse.
    Fitting the data simultaneously with these two equations yields the bright-state lifetime $\tau_b = \SI{110(8)}{\micro\second}$ and dark-state lifetime $\tau_d= \SI{37(2)}{\micro\second}$, see~\figref{fig:lifetime}{b}. 
    With a suitable combination of repumpers in the read-out stage, we can further identify the branching into each substate among the bright states, see~\figref{fig:lifetime}{c}. 
    Specifically, we measure the population accumulation rate of the bright state $\gamma_b$, and the same quantity after (i).\ 461\,nm (ii).\ 496\,nm+461\,nm pulses. 
    Based on their ratio, the branching can be calculated.
    Note that the observed relative branching within the bright states is consistent with the expectation considering only the degeneracy alone.

    \section{Rydberg coherence}
\begin{figure*}[t]
        \centering
        \includegraphics[scale=1]{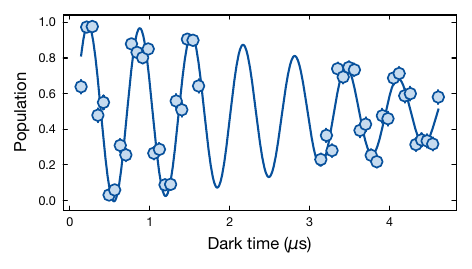}
        \caption{
        \textbf{Rydberg coherence time.}
        The envelope function $C(t) = \exp\left( -2 \pi^2 \sigma^2 t^2 \right)$ used in the fit assumes a detuning drift following the Gaussian distribution. 
        The coherence time $T_2^* = \SI{4.3(4)}{\micro\second}$.
       }
        \label{fig:coherence}
    \end{figure*}

    We measured the coherence time ($T_2^*$) on the $\ket 1$-$\ket r$ transition using a Ramsey experiment, see~\figref{fig:coherence}{}. 
    We anticipate the slow detuning drift as the primary source of decoherence and fit the data using the envelope function $C(t) = \exp\left( -2 \pi^2 \sigma^2 t^2 \right)$. 
    This expression can be derived by averaging Ramsey traces over detunings randomly sampled from a Gaussian distribution of width $\sigma$. 
    The fitted $\sigma$ is about $53\,$kHz, which corresponds to $T_2^* = \SI{4.3(4)}{\micro\second}$.

    \section{Two-qubit benchmarking sequence}
    During the entangling gates, we rapidly turn off the tweezer traps.
    The entangling gates are then carried out in free-space, to avoid effects such as anti-trapping when atoms are in the Rydberg state. 

    To optimize the gate parameters, we use the interleaved echo sequence presented in Ref.~\cite{Evered2023} for its insensitivity to single-qubit phase. 
    The objective of our optimization is to maximize the return probability of $\ket {11}$ state after applying a constant number (10 in our case) of CZ gates on initial state $\ket {11}$. 
    The parameters of the phase modulation are scanned experimentally using a numerically calculated Hessian~\cite{Muniz2024}, which is derived based on an estimated AOM rise-time and Rydberg interaction shift ($2\pi \times 114\,$MHz). 
    Usually, a good convergence can be reached in one round of optimization. 
    For quoting the gate fidelity, we use the recently presented symmetric stabilizer benchmarking sequence (SSB)~\cite{Tsai2025}, which we find to be more robust than the interleaved echo sequence.
    %
    %
    Once a set of optimal gate parameters are found, we apply them in a SSB sequence, and scan the single-qubit phase to maximize $\ket{11}$ return probability. 
    %



    \begin{figure*}[t]
        \centering
        \includegraphics{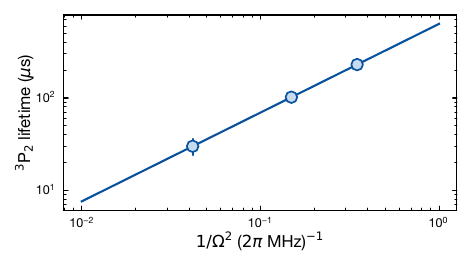}
        \caption{
        \textbf{${}^3\text{P}_2$ loss in the presence of 316 nm light.}
        ${}^3\text{P}_2$ lifetime $\tau$ scales with UV rabi frequency $\Omega$ as $\tau = A/\Omega^2$ with a proportionality constant $A=610/\SI{}{\micro\second}$.
       }
        \label{fig:ionization_loss}
    \end{figure*}

    \section{Error budget}
    We estimate the contributions to the infidelity of our gate using the effective super-operator associated with each error source. 
    The super-operator is obtained, for each error source, by solving the associated master equation, taking into account the coherent error sources in the Hamiltonian and the collapse operators for spontaneous decay, pure dephasing on the Rydberg transition, and $^3\text{P}_2$ ionization.
    The infidelity is calculated in two ways, through the process fidelity between the simulated super-operator with the ideal one, and using a simulated SSB sequence with the obtained super-operator. 
    We observe no notable difference in results between the two methods. 
    For gate infidelity originating from laser phase noise, we compared our result to the one derived from the recently proposed linear response calculation~\cite{Tsai2025} and find good agreement. 
    Numerically, aside from the qubit and Rydberg states, we define an additional ``bucket" state for each atom to which it is assumed to be transferred in the case of either Rydberg decay to outside the qubit manifold or ${}^3\text{P}_2$ ionization loss, see~\figref{fig:ionization_loss}{}.

\begin{figure*}[t]
    \centering
    \includegraphics[scale=1]{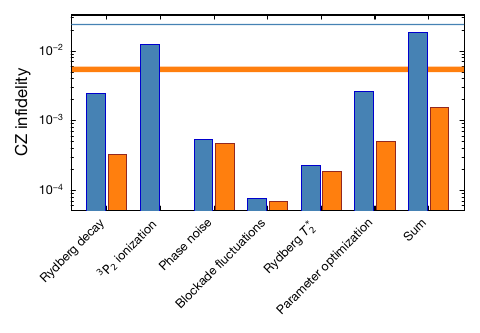}
    \caption{
    \textbf{CZ gate error budget.} 
    Leading contributions to the infidelity of the CZ gate with (orange) and without (blue) loss correction at a Rabi frequency of $2\pi \times 6\,$ MHz.
    The experimental infidelity are indicated by the corresponding horizontal lines. 
   }
    \label{fig:errors}
\end{figure*}

    To estimate infidelities taking the loss-correcting SRD into account, we use the simulated SSB sequence once more, but only consider the valid computational states. 
    Population decayed from Rydberg states into $m_J \neq 0$ states of $^3\text{P}_2$ which are not qubit states but nevertheless manifest as if they are, are also taken into account. 
    We identify Rydberg state decay, $^3\text{P}_2$ ionization (see above), and gate parameter estimation errors as the most prominent, known, error sources in our two-qubit gate. 
    The effect of other known noise sources are presented in \figref{fig:errors}{}. 
    The modeled noise sources give rise to a total raw infidelity of $1.84\%$ and loss-corrected infidelity of $0.16\%$. 
    We attribute the higher infidelities measured in the experiment to experimental drifts such as in Rabi frequency and UV pulse area, phase, and shape.
    Note that the ionization loss will be eliminated with a UV laser system coupling ${}^3\text{P}_2$ to the Rydberg state, and the infidelity from Rydberg decay can be further reduced by increasing the laser power.
 
\bibliography{reference}

\end{document}